\documentclass[utf8]{frontiersSCNS} % for Science, Engineering and Humanities and Social Sciences articles
\usepackage{url,hyperref,lineno,microtype,subcaption}
\usepackage[onehalfspacing]{setspace}
\usepackage{hyperref}
\def\keyFont{\fontsize{8}{11}\helveticabold }
\def\firstAuthorLast{Li {et~al.}} %use et al only if is more than 1 author
\def\Authors{Dong~Li\,$^{1,*}$, Fanpeng~Shi\,$^{1,2}$, Haisheng~Zhao\,$^{3}$, Shaolin~Xiong\,$^{3}$, Liming~Song\,$^{3}$, Wenxi~Peng\,$^{3}$, Xinqiao~Li\,$^{3}$, Wei~Chen\,$^{1}$ and Zongjun~Ning\,$^{1,2}$}
\def\Address{$^{1}$Purple Mountain Observatory, Chinese Academy of Sciences, Nanjing 210023, China \\
$^{2}$School of Astronomy and Space Science, University of Science and Technology of China, Hefei 230026, China \\
$^{3}$Key Laboratory of Particle Astrophysics, Institute of High Energy Physics, Chinese Academy of Sciences, Beijing 100049, China }
\def\corrAuthor{Dong Li}

\def\corrEmail{lidong@pmo.ac.cn}

\begin{document}
\onecolumn
\firstpage{1}

\title[Flare Quasi-Periodic Pulsation Associated with Recurrent Jets]{Flare Quasi-Periodic Pulsation Associated with Recurrent Jets}

\author[\firstAuthorLast ]{\Authors} %This field will be automatically populated
\address{} %This field will be automatically populated
\correspondance{} %This field will be automatically populated
\extraAuth{}% If there are more than 1 corresponding author, comment this line and uncomment the next one.

\maketitle
%%%%%%%%%%%%%%%%%%%%%%%%%%%%%%%%%%%%%%%%%%%%%%%%%%%%%%%%%%%%%%%%%%%%%%%%%%%%%%%%%%%%%%%%%%%%%%%%%%%
\begin{abstract}

\section{}
Quasi-periodic pulsations (QPPs), which carry time features and
plasma characteristics of flare emissions, are frequently observed
in light curves of solar/stellar flares. In this paper, we
investigate non-stationary QPPs associated with recurrent jets during
an M1.2 flare on 2022 July 14. A quasi-period of $\sim$45$\pm$10~s,
determined by the wavelet transform technique, is simultaneously
identified at wavelengths of soft/hard X-ray and microwave
emissions, which are recorded by the Gravitational wave high-energy
Electromagnetic Counterpart All-sky Monitor, Fermi, and the Nobeyama
Radio Polarimeters, respectively. A group of recurrent jets with an
intermittent cadence of about 45$\pm$10~s are found in Atmospheric
Imaging Assembly (AIA) image series at 304~{\AA}, but they are 180-s
earlier than the flare QPP. All observational facts suggest that the
flare QPP could be excited by recurrent jets, and they should be
associated with nonthermal electrons that are periodically
accelerated by a repeated energy release process, like repetitive
magnetic reconnection. Moreover, the same quasi-period is discovered
at double footpoints connected by a hot flare loop in AIA~94~{\AA},
and the phase speed is measured to $\sim$1420~km~s$^{-1}$. Based on
the differential emission measure, the average temperatures, number
densities, and magnetic field strengths at the loop top and
footpoint are estimated to $\sim$7.7/6.7~MK,
$\sim$7.5/3.6$\times$10$^{10}$~cm$^{-3}$, and $\sim$143/99~G,
respectively. Our measurements indicate that the 45-s QPP is
probably modulated by the kink-mode wave of the flare loop.

\tiny
 \keyFont{\section{Keywords:} Sun: flares, Sun: oscillations, Sun: UV emission, Sun: X-ray
emission, Sun: radio emission, MHD waves}
\end{abstract}

\section{Introduction}
Quasi-periodic pulsations (QPPs) observed in solar/stellar flares
usually appear as temporal intensity oscillations of electromagnetic
radiation \citep[see,][and references therein]{Kupriyanova20,
Zimovets21}. They are frequently identified as a series of
repetitive but irregular pulsations with anharmonic and symmetric
triangular shapes, referring to non-stationary QPPs
\citep[e.g.,][]{Nakariakov19}. The observation of QPPs has been
reported in flare time series over a broad range of wavelengths,
ranging from radio/microwave emissions through ultraviolet (UV) and
white light wavelengths to soft and hard X-rays (SXR/HXR) channels,
and even in the $\gamma$-ray emission
\citep[e.g.,][]{Nakariakov10a,Tan16,Milligan17,Li17a,Kashapova21,Kolotkov21,Lu21,Doyle22,Smith22,Zhang22b}.
Generally, a typical QPP should be at least three successive and
complete pulsations. There is not reason to talk about the QPP
behavior if there are only one or two pulsations, which might be
just a coincidence, for instance, the similar time interval between
successive pulsations occurred by chance \citep{Nakariakov19}. The
characteristic time of all pulsations in one QPP event is expected
to be same, which can be regarded as the period. However, the
characteristic time of these pulsations could be varied, indicating
the irregular nature of flare QPPs. Thus, they often show the
variation of quasi-periods \citep[e.g.,][]{Nakariakov18}. In
observations, the quasi-periods of flare QPPs are found to vary from
a fraction of seconds to a few dozens of minutes
\citep{Tan10,Yuan13,Ning14,Meszarosova16,Kolotkov18,Hayes20,Karlicky20,Hong21,Bate22}.

It has been accepted that the quasi-periods of flare QPPs are often
related to their generation mechanisms \citep{Kupriyanova20}. The
short-period (i.e., $<$1~s) QPPs, which are usually observed in
radio/microwave emissions, are often driven by the dynamic
interaction between plasma waves and energetic particles in complex
magnetic structures \citep{Nakariakov18,Yu19,Karlicky20}. The flare
QPPs with long periods in the order of seconds and minutes, which
could detect in almost all wavelengths, are frequently interpreted
in terms of magnetohydrodynamic (MHD) waves in slow modes
\citep[e.g.,][]{Wang21}, kink modes \citep[e.g.,][]{Nakariakov21},
and sausage modes \citep[e.g.,][]{Lib20}. In such case, the flare
QPPs with periods larger than 1~minute could be associated with slow
sausage waves \citep{Sadeghi19,Gao21}, global kink waves
\citep{Duckenfield19,Gao22}, and slow magnetoacoustic waves
\citep{Wang11,Ofman12,Yuan15,Prasad22}; while those with periods in
the order of seconds are often explained as fast sausage or kink
waves \citep{Inglis09,Guo21,Kashapova21}, depending on whether the
plasma loop can be compressible or incompressible
\citep{Yuan16,Nakariakov20}. Those long-period QPPs might be also
associated with the repetitive magnetic reconnection
\citep{Thurgood19,Karampelas22}. The idea is that the released
energy via intermittent magnetic reconnection is repeated, which can
periodically accelerate nonthermal electrons. Thus, it is often used
to explain the QPPs seen in the impulsive phase of solar flares
\citep[e.g.,][]{Yuan19,Li21}. Moreover, this reconnection process
could either be spontaneous such as `magnetic dripping'
\citep[e.g.,][]{Nakariakov10} and `magnetic tuning fork'
\citep[e.g.,][]{Takasao16}, or it might be triggered by an external
MHD wave \citep{Foullon05,Nakariakov18}.

Solar jets, which often show columnar and beam-like
structures, are usually associated with solar flares, type III radio
bursts, and filament eruptions
\citep{Shibata07,Shen11,Paraschiv15,Raouafi16}. They can be observed
everywhere on the Sun, such as active regions, quiet-Sun regions,
and coronal holes \citep{Brueckner83,Shen21}. The recurrent jets,
which always reveal ejected plasmas repeatedly and have the same
base source \citep{Tian18,Lu19}, become a topic of particular
interest because they could be associated with flare QPPs
\citep{Ning22,Shi22}, fast-mode EUV waves and quasi-periodic
fast-propagating (QFP) magnetosonic waves \citep{Shen18a,Shen18b}.
The observed QFP waves often consist of multiple concentric and
coherent wavefronts, termed as `QFP wave trains', and they are
produced successively within periods of dozens of seconds or a few
minutes near the epicenter of the accompanying flares
\citep{Shen22,Shen22b}. Sometimes, the quasi-periods of QFP wave
trains are quite similar to those of associated flare QPPs, implying
that the two different phenomena might manifest the two different
aspects of the same physical process, i.e., the pulsed energy
release via repeating magnetic reconnection
\citep{Liu11,Shen12b,Shen13,Shen18,Kolotkov18,Zhou22}. On the other
hand, some quasi-periods of QFP wave trains are completely
unassociated with those of flares QPPs, indicating that the
periodicity of QFP wave trains is diverse and could not be
associated with flare QPPs \citep{Shen18c,Shen19}. Therefore, the
relationship between flare QPPs and QFP wave trains still needs
in-depth investigation \citep{Shen22}.

The observed QPPs could provide the time feature and plasma
characteristic of flare emissions, which are helpful for diagnosing
plasma properties on the Sun or Sun-like stars, especially at the
flare location \citep{Pugh19,Zimovets21}. When considering that
flare QPPs are modulated by MHD waves, they might potentially lead
to coronal heating through dissipating of those waves
\citep{Reale19,Van20,White21,Li22a}. Moreover, they can allow us to
map coronal magnetic fields and estimate plasma parameters in the
corona, named as `coronal seismology'
\citep[e.g.,][]{Yang20,Anfinogentov22}. In this paper, we report
multi-wavelength observations of the flare QPP associated with
recurrent jets, and the flare QPP is also found at two opposite
footpoints connected by a hot flare loop seen in AIA~94~{\AA}
images. Our measurements suggest that the flare QPP could be
interpreted as kink-mode MHD wave of the flare loop.

\section{Observations}
\begin{figure}[h!]
\begin{center}
\includegraphics[width=0.8\linewidth]{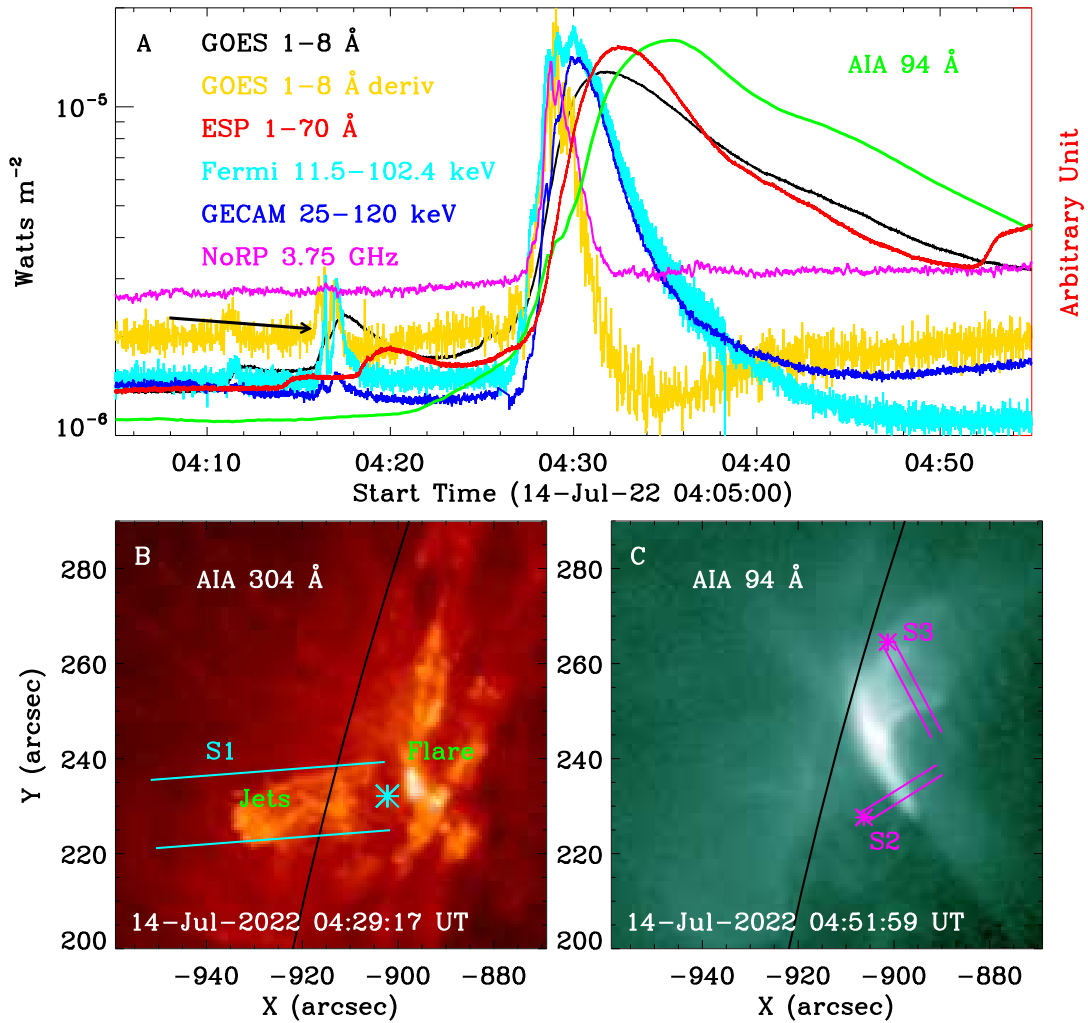}
\end{center}
\caption{Overview of the solar flare on 2022 July 14. (A) Full-disk
light curves from 04:05~UT to 04:55~UT recoded by GOES (black),
EVE/ESP (red), Fermi/GBM (cyan), GECAM (blue), and NoRP (magenta).
The local light curve in AIA~94~{\AA} (green), which is integrated
over the flare region in panel~(C). (B-C) Snapshot with a FOV of
90$''$$\times$90$''$ captured by SDO/AIA at wavelengths of 304~{\AA}
and 94~{\AA}. Two cyan lines outline the slit (S1) that contains
the solar jets, and the magenta lines mark two slits (S2 and S3)
across double footpoints. The color symbols of `$\ast$' indicate
their beginning points, and the black curve represents the solar
limb.} \label{over}
\end{figure}

On 2022 July 14, a solar flare occurred in the active region NOAA
13058 (N15E81), which was close to the solar limb and erupted after
a group of recurrent jets. It was simultaneously
observed by several space-based telescopes, such as the
Geostationary Operational Environmental Satellite X-ray Sensor
\citep[GOES/XRS;][]{Hanser96}, the Fermi Gamma-ray Burst Monitor
\citep[GBM;][]{Meegan09}, the Gravitational wave high-energy
Electromagnetic Counterpart All-sky Monitor
\citep[GECAM;][]{Xiao22}, the Atmospheric Imaging Assembly
\citep[AIA;][]{Lemen12} and the Extreme Ultraviolet Variability
Experiment \citep[EVE;][]{Woods12} on board the Solar Dynamics
Observatory \citep[SDO;][]{Pesnell12}, and the ground-based radio
telescope, i.e., the Nobeyama Radio Polarimeters
\citep[NoRP;][]{Nakajima85}, as seen in Table~\ref{tab1} and
Figure~\ref{over}. It should be pointed out that all light curves
expected for GOES have been multiplied by a factor, so that they can
be well displayed in a same window.

GOES/XRS \citep{Hanser96,Lotoaniu17} is used to monitor the
full-disk solar irradiance at SXR channels with a time cadence of
1~s, particularly for monitoring the flare emission, as shown by the
black line in Figure~\ref{over}~(A). According to the
GOES~1$-$8~{\AA} flux, the solar flare was identified as an M1.2
class, it began at $\sim$04:22~UT, reached its maximum at about
04:31~UT, and stopped at $\sim$04:40~UT. The gold line shows
the derivative flux at GOES~1$-$8~{\AA}. The EUV SpectroPhotometer
\citep[ESP;][]{Didkovsky12} for SDO/EVE could also provide the SXR
flux at 1$-$70~{\AA} with a time cadence of 0.25~s, as indicated by
the red line. The SXR light curves observed by GOES and ESP match
well with each other, and they both appear double peaks before the
onset time of the M1.2 flare, i.e., from 04:16~UT to 04:20~UT, as
indicated by the black arrow. They might be a candidate of the flare
precursor.

Fermi/GBM can provide the solar irradiance that is integrated over
the whole Sun at both SXR and HXR channels. The temporal cadence is
commonly 0.256~s, but it becomes 0.064~s automatically during solar
flares \citep{Meegan09}. Thus, we first interpolate them into an
uniform temporal resolution of 0.256~s before analysing and such
temporal resolution is sufficient to study the flare QPP with a
quasi-period of tens of seconds \citep[cf.][]{Li15,Ning17}.
Figure~\ref{over}~(A) draws the Fermi/GBM light curve at
11.5$-$102.4~keV, as shown by the cyan curve, which is measured by
the n5 detector. GECAM is designed to detect and localize
high-energy transients, such as Gamma-ray bursts and solar flares.
It consists of 25 gamma-ray detectors (GRDs), which are used to
detect the X-ray and $\gamma$-ray radiation \citep{Xiao22}.
Figure~\ref{over}~(A) shows the solar flux at 25$-$120~keV (blue)
during the M1.2 flare with an uniform temporal cadence of 0.5~s, the
GRD numbers and their averaged incident angles for each GRD used in
this study are listed in Table~\ref{tab2}. Both Fermi/GBM and
GECAM/GRD light curves appear double peaks between about 04:16~UT
and 04:20~UT, similarly to what have seen in SXR fluxes recorded by
GOES/XRS and SDO/EVE/ESP.

The M1.2 flare was also observed by NoRP at the radio/microwave
emission with a temporal cadence of 1~s, as shown by the magenta
line in Figure~\ref{over}~(A). It matches well with the
GOES~1-8~{\AA} derivative flux, indicating the Neupert effect during
the M1.2 flare \citep[cf.][]{Neupert68}. The microwave flux also
reveals several successive sub-peaks during the flare impulsive
phase, similarly to what observed in the Fermi (cyan) and GECAM
(blue) light curves, which could be regarded as QPPs. On the other
hand, we do not see the small peak before the M1.2 flare in the NoRP
light curve. So, it is impossible to determine a flare precursor
here. Fortunately, SDO/AIA can provide full-disk spatial-resolved
maps in seven EUV and two UV wavelength bands. The spatial scale for
each AIA map is 0.6$''$~per~pixel, and the temporal cadence is 12~s
for EUV maps. Before analysing, all the AIA maps have been
preprocessed by `aia\_prep.pro' \citep{Lemen12}.
Figure~\ref{over}~(B-C) presents AIA maps with a sub-field of about
90$''$$\times$90$''$ at 304~{\AA} and 94~{\AA}, respectively. A
group of jets can be seen in the AIA~304~{\AA} map (see also the
animation.mp4), as outlined by two cyan lines. In order to cover the
bulk of these jets as much as possible during their lifetime, we
used a constant width of about 15$''$. The base of jets is close to
one flare ribbon in AIA~304~{\AA} maps. A post flare loop can be
seen in the AIA~94~{\AA} map, and two pairs of magenta lines with a
width of about 3$''$ are used to outline double footpoints (or loop
legs). Finally, the light curve at AIA~94~{\AA} is integrated from
the flare region, as indicated by the green line in
Figure~\ref{over}~(A). We can not see the small peak during
$\sim$04:16$-$04:20~UT before the flare onset. Thus, we can conclude
that the small double peaks (indicated by the black arrow) in
SXR/HXR channels are not identified as the flare precursor
\citep[e.g.,][]{Dudk16,Benz17,Yan17,Li18b,Li20c}

\section{Results and Discussions}
\subsection{Multi-wavelength observations of flare QPP}
\begin{figure}[h!]
\begin{center}
\includegraphics[width=0.8\linewidth]{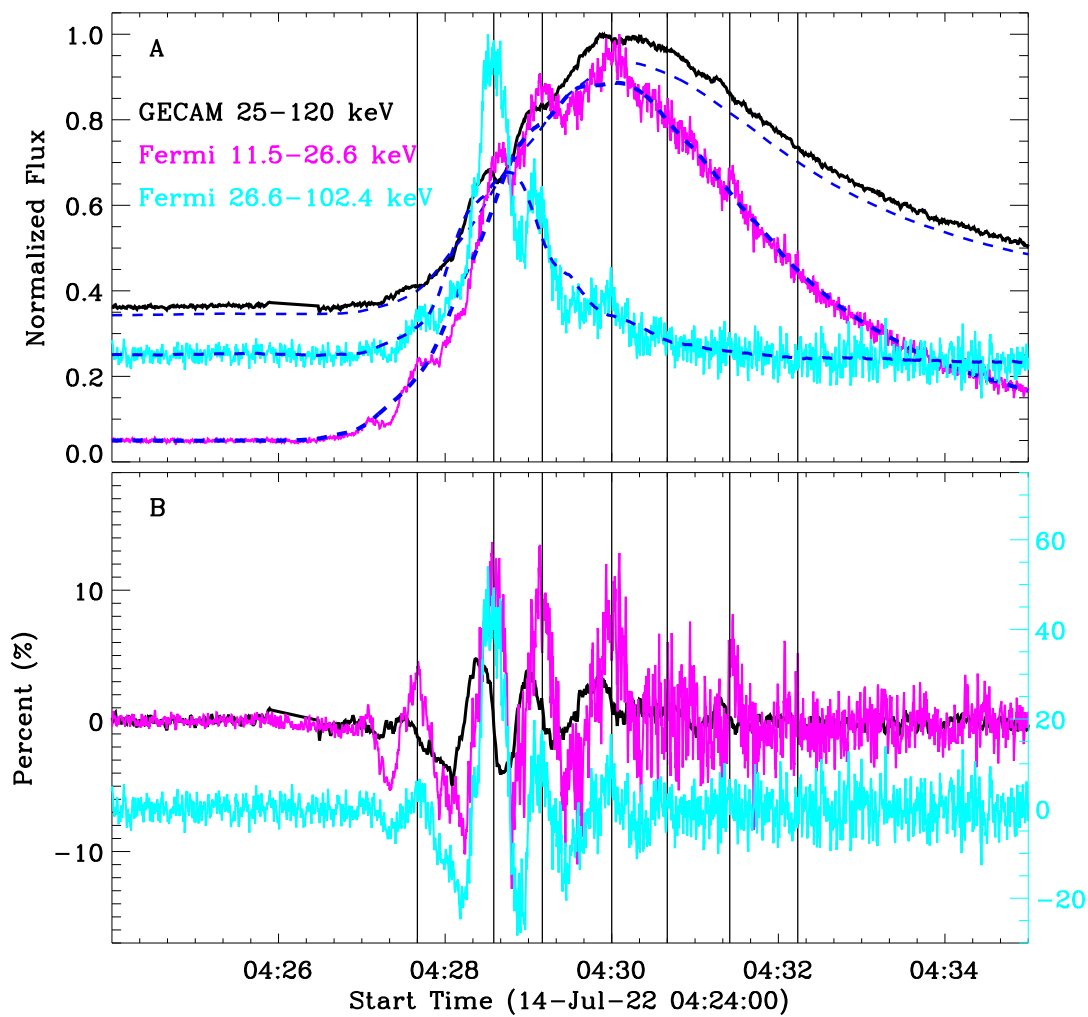}
\end{center}
\caption{X-ray light curves of the M1.2 flare. (A) Normalized
HXR/SXR fluxes recorded by GECAM (black) and Fermi/GBM (magenta and
cyan), the overlaid dashed lines are their slowly varying
components. (B) The corresponding rapidly varying components that
are normalized to their maximum slow-varying components. The
vertical lines indicate HXR peaks during the solar flare.}
\label{flux}
\end{figure}

The small double peaks before the M1.2 flare seen in SXR/HXR fluxes
can not be regarded as the flare precursor, because they are not
homologous with the flare source, as shown in Figure~\ref{over}.
Herein, only the successive sub-peaks seen in HXR and microwave
emissions during the flare impulsive (i.e., $\sim$04:27$-$04:32~UT)
are investigated in this study. Figure~\ref{flux} presents HXR light
curves at GECAM~25$-$120~keV (black), Fermi~11.5$-$26.6~keV
(magenta) and 26.6$-$102.4~keV (cyan). They appear to be
characterized by several small-amplitude sub-peaks superimposed on
the large-amplitude pulse. These sub-peaks with small amplitudes are
successive and could be regarded as QPPs, while the main pulse with
the large amplitude can be regraded as a strong background trend.
The vertical lines indicate seven sub-peaks from roughly 04:27:50~UT
to about 04:32:20~UT, and the average duration is 45~s,
corresponding to a quasi-period of 45~s. We also note that some
sub-peaks might be not very clear in the raw light curve, largely
due to their small amplitudes. Using a smooth window of 60~s
\citep{Nakariakov10a,Yuan11,Li15,Li22}, the raw light curve is
decomposed into two components: a rapidly varying component (QPP)
plus a slowly varying component (background). Thereby, the
shorter-period oscillation (i.e, 45-s QPP) is enhanced, while the
long-period background trend is suppressed \citep[see][for the
discussion of this method]{Kupriyanova10,Gruber11,Auchere16}. The
overplotted blue dashed lines represent the slowly varying
components, and the rapidly varying components are shown in
panel~(B). Obviously, the rapidly varying components are dominated
by the QPP feature, i.e., some repetitive but irregular pulsations,
as marked by the vertical lines. They match well with the successive
sub-peaks seen in the raw light curves, indicating that the smooth
method only enhance the short-period oscillation, but does not
change it. Therefore, these repetitive but irregular pulsations
could regard as the signature of non-stationary QPPs
\citep[cf.][]{Nakariakov19}, and they can not be the artifact of
smoothing \citep[cf.][]{Li21}. Here, the modulation depth of flare
QPPs, which is regarded as the ratio between rapidly varying
components and the maximum value of slowly varying components, are
roughly equal to 10\%$-$25\%. This result is consistent with
previous findings for flare QPPs in HXR emissions
\citep[e.g.,][]{Nakariakov10a,Li22}.

\begin{figure}[h!]
\begin{center}
\includegraphics[width=0.8\linewidth]{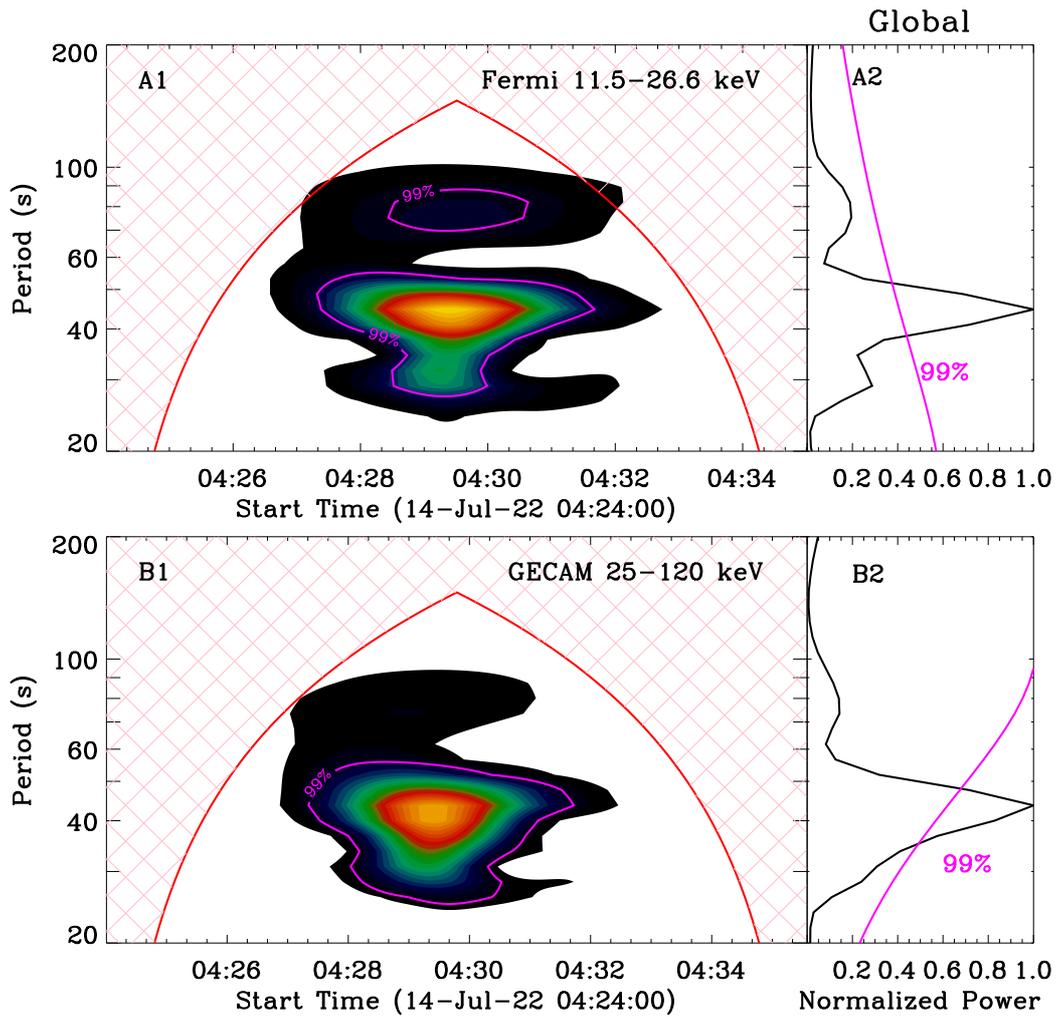}
\end{center}
\caption{Wavelet analysis results of the rapidly varying components
in Figure~\ref{flux}. (A1-A2) The wavelet power spectrum and global
wavelet power at Fermi~11.5$-$26.6~keV. (B1-B2) The wavelet power
spectrum and global wavelet power at GECAM~25$-$120~keV. The magenta
lines indicate the significance level of 99\%, and the red curve
outline a confidence interval.} \label{hxr}
\end{figure}

Next, the Morlet wavelet analysis method is applied to the rapidly
varying components at Fermi~11.5$-$26.6~keV and GECAM~25$-$120~keV,
as shown in Figure~\ref{hxr}. Based on the Parseval's theorem for
wavelet analysis \citep{Torrence98}, the wavelet power has been
normalized, which could provide the conservation of total energy
signals under the wavelet transform, and then obtained a
distribution of the spectral power across wavelet periods.
Panels~(A1) and (B1) show the wavelet power spectra, and they both
exhibit an enhanced power over a wide range in almost the same time
interval from about 04:27:50~UT to 04:32:20~UT, indicating a flare
QPP within large uncertainties. The bulk of power spectrum (at the
confidence level of 99\%) is dominated by a quasi-period centered at
$\sim$45~s. The dominant period of $\sim$45~s is confirmed by the
global wavelet power spectrum, as shown in panels~(A2) and (B2).
From which, a significant peak at about 45~s is seen in the global
wavelet power spectrum. On the other hand, the period uncertainty of
$\pm$10~s could be determined by the full width at half maximum
value of the peak global power above the 99\% confidence level
\citep[as performed by][]{Yuan11,Tian16,Li20a}.

\begin{figure}[h!]
\begin{center}
\includegraphics[width=0.8\linewidth]{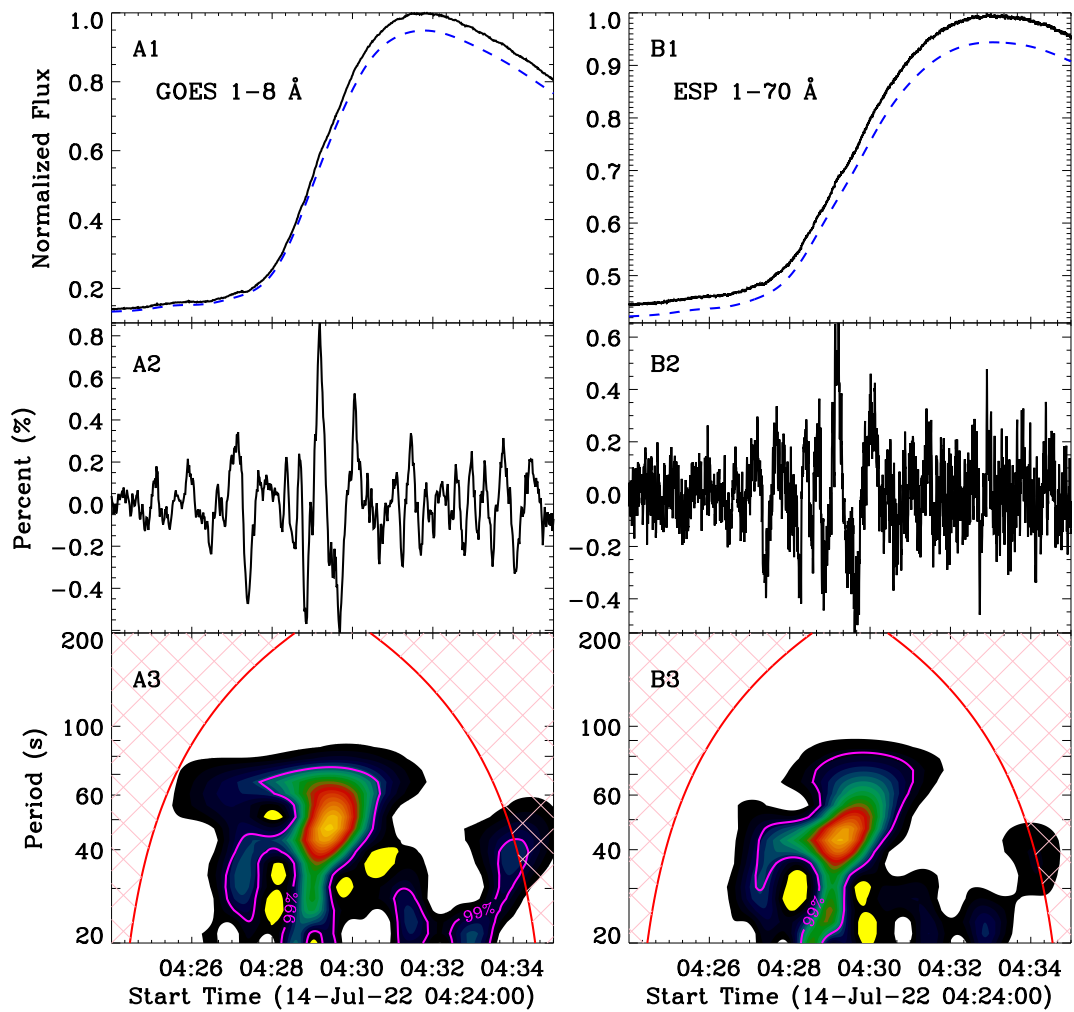}
\end{center}
\caption{Morlet wavelet analysis results in SXR wavelengths. (A1-B1)
Normalized SXR fluxes observed by GOES~1$-$8~{\AA} and
ESP~1$-$70~{\AA}, the overlaid dashed lines are their slowly varying
components, after multiplication by 0.95. (A2-B2) The corresponding
rapidly varying components, which are normalized to their maximum
slow-varying components. (A3-B3) Morlet wavelet power spectra of the
rapidly varying components. The magenta lines indicate the
significance level of 99\%.} \label{sxr}
\end{figure}

The flare QPP with a quasi-period of about 45$\pm$10~s is seen in
the HXR radiation observed by Fermi and GECAM. However, the Fermi
flux at 11.5$-$26.6~keV might consist of SXR and HXR components. In
order to know if the flare QPP could be found in the SXR emission,
we then perform the Morlet wavelet analysis on SXR light curves at
GOES~1$-$8~{\AA} and ESP~1$-$70~{\AA}, as shown in Figure~\ref{sxr}.
Panels~(A1) and (B1) present the raw SXR light curves (black) and
their slowly varying components (dashed blue) after applying a
smooth window of 60~s. It should be pointed out that the slowly
varying components have been multiplied by 0.95 to avoid overlap
with the raw light curves \citep[cf.][]{Ning22}. Panels~(A2) and
(B2) plot the corresponding rapidly varying components, which are
characterized by a series of successive pulsations. The modulation
depth of SXR radiation is only about 0.4\%$-$0.6\%, which is much
smaller than that of HXR emissions. This is consistent with previous
observations, for instance, the flare SXR emission often reveals the
small-amplitude oscillation, while the HXR QPP usually has a large
amplitude \citep[e.g.,][]{Nakariakov10a,Ning17,Li20b,Ning22}.
Panels~(A3) and (B3) show the Morlet wavelet power spectra of
rapidly varying components. They both reveal an enhanced power at
the period center of about 45~s over a time interval from roughly
04:27~UT to 04:31~UT, suggesting a dominant period of $\sim$45~s,
similarly to what observed in HXR QPPs.

\begin{figure}[h!]
\begin{center}
\includegraphics[width=0.8\linewidth]{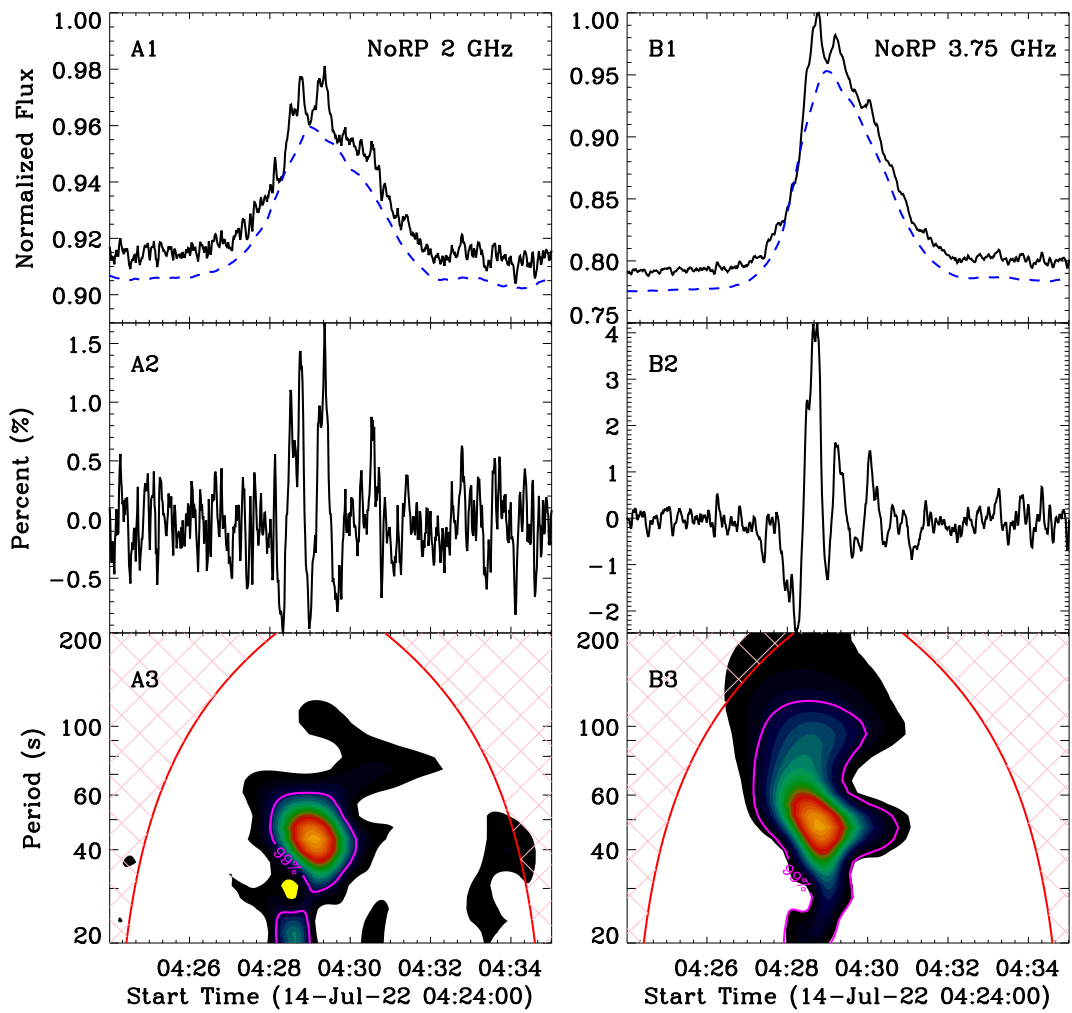}
\end{center}
\caption{Similar to Figure~\ref{sxr}, but the Morlet wavelet
analysis is performed for radio fluxes at frequencies of NoRP~2~GHz
and 3.75~GHz.} \label{radio}
\end{figure}

Figure~\ref{radio} presents the Morlet wavelet analysis on radio
fluxes at frequencies of NoRP~2~GHz (A1-A3) and 3.75~GHz (B1-B3).
Using the same smooth window of 60~s, the raw light curves (black)
are decomposed into slowly (dashed blue) and rapidly varying
components (A2-B2). The modulation depth of radio QPPs is estimated
to about 1\%$-$2\%, which is larger than that of SXR QPPs, but is
still smaller than that of HXR QPPs. We also note that only 3 or 4
successive pulsations appear in radio fluxes, which are less than
that in HXR fluxes. On the other hand, a same quasi-period centered
at $\sim$45~s is seen in the wavelet power spectrum, which agrees
with the 45-s QPP observed by Fermi, GECAM, GOES and EVE/ESP.
The same quasi-period of 45~s is simultaneously detected in
SXR, HXR, and microwave emissions during the impulsive phase of the
M1.2 flare, suggesting that the 45-s QPP seen at multiple
wavelengths should originate from a same process of energy release,
i.e., the repetitive magnetic reconnection.

The flare QPP could be observed at multiple wavelengths of HXR, SXR,
and microwave emissions, suggesting that the 45-s QPP
simultaneously appear in both the nonthermal and thermal emissions.
In other words, the nonthermal and thermal processes could be
coexisted during the M1.2 flare
\citep[e.g.,][]{Warmuth16,Li20a,Ning22}. The 45-s QPP observed in
the thermal emission at SXR wavelengths may share the same origin as
the QPP feature seen in the nonthermal emission at HXR and microwave
channels. The M1.2 flare showed the Neupert effect
(Figire~\ref{over}), which is a plasma heating via energy releasing
through electron beams \citep{Neupert68,Ning08,Ning09}. The flare
QPP observed at multiple wavelengths is most likely to be associated
with the nonthermal process, i.e., the periodically accelerated
electron beams via the repetitive magnetic reconnection
\citep[e.g.,][]{Li21,Karampelas22}. The idea is that the released
energy via periodic reconnection could periodically accelerate
electron beams, producing repetitive HXR and microwave pulsations in
the solar corona. Meanwhile, the repeated SXR pulsations are
periodically generated by plasma heating after magnetic reconnection
\citep[see][for a recent review]{Zimovets21}.

\subsection{Recurrent jets associated with flare QPP}
\begin{figure}[h!]
\begin{center}
\includegraphics[width=0.8\linewidth]{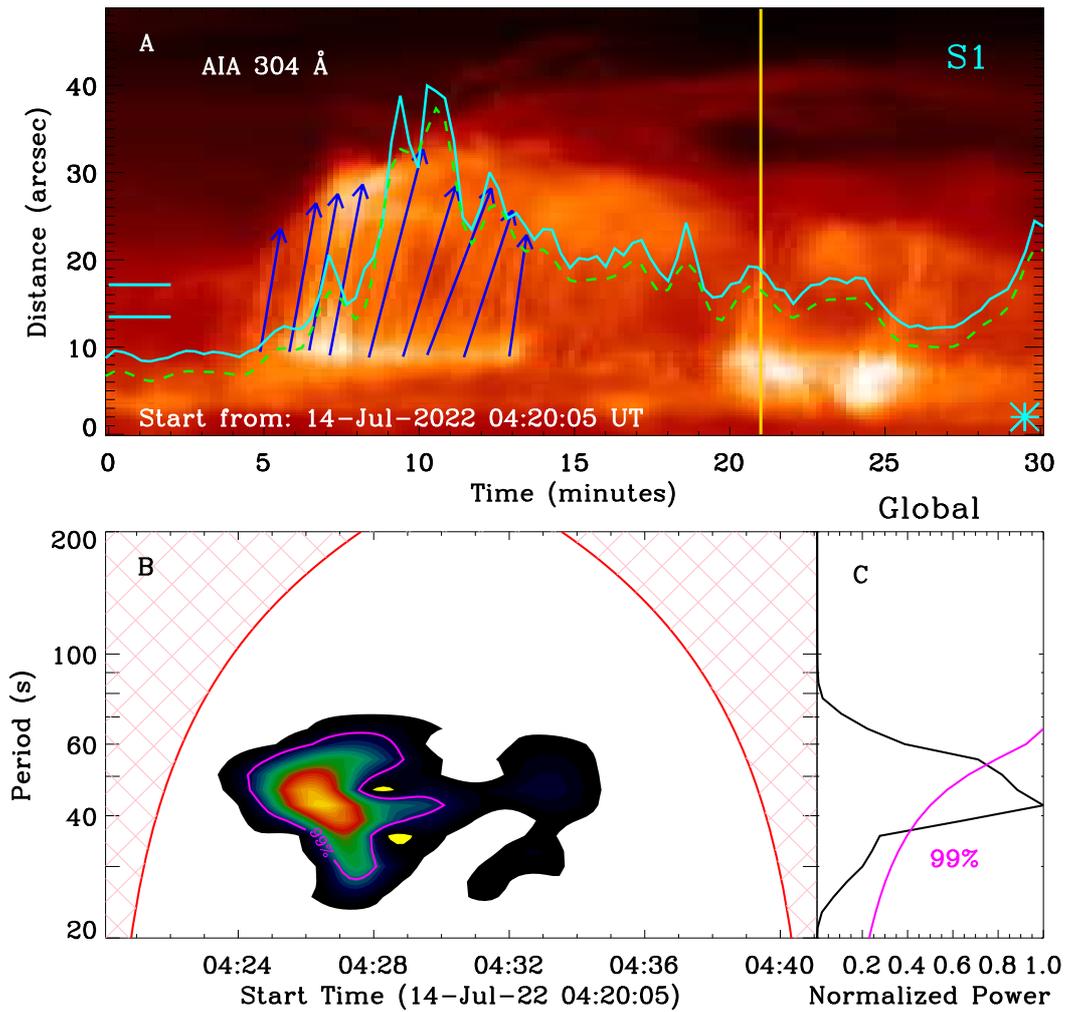}
\end{center}
\caption{Solar jets observed by SDO/AIA~304~{\AA}. (A) Time-distance
diagram along S1 (Figure~\ref{over}), the cyan symbol ($\ast$) marks
the start point. The overlaid solid cyan curve is integrated
from two short cyan lines on the left hand, and the green
dashed curve represents its slowly varying component. The blue
arrows mark the recurrent jets. The vertical gold
line mark the stop time in panel~(B). (B-C) Morlet wavelet power
spectrum and global wavelet power. The magenta lines indicate the
significance level of 99\%.} \label{jet}
\end{figure}

Figure~\ref{over}~(B) and the animation.mp4 show that a group of
plasma ejections during the M1.2 flare. They manifest as
collimated and beam-like structures in AIA~304~{\AA}, which could be
identified as `solar jets' \citep[e.g.,][]{Shen21}. To look closely
the jet eruptions and periodicity, we draw the time-distance image
along the slit of S1 that is made from AIA~304~{\AA} image series,
as shown in Figure~\ref{jet}~(A). Here, the slit is selected to be a
constant width of about 15$''$, and thus it can cover the bulk of
jet bodies as much as possible. A series of solar jets can be seen
in the time-distance image, and their apparent speed is estimated to
about 110$-$300~km~s$^{-1}$, as indicated by the blue arrows. A
total of nine jets are found during the time interval of about
450~s, and the average intermittent cadence is roughly equal to
50~s. Such intermittent cadence is quite close to the quasi-period
of the flare QPP, implying that those jets occur periodically. Then
the intensity variation integrated over two short cyan lines is
overplotted, as shown by the cyan line. The intensity curve seems to
reveal several sub-peaks corresponding to solar jets. However, it is
hard to show a one-to-one correspondence, mainly due to the
small-amplitude sub-peaks superimposed on the strong background
emission. Therefore, the slowly (dashed green) and rapidly varying
components are distinguished with the smooth window of 60~s, and the
Morlet wavelet analysis is applied to the rapidly varying component.
Panels~(B) and (C) show the Morlet wavelet power spectrum and its
global wavelet power spectrum. They both reveal a period centered at
about 45~s, confirming that the recurrent jets are associated with
flare QPPs. Moreover, the recurrent jets appear to start at about
04:24:50~UT, which are $\sim$180-s earlier than the flare QPP. Our
observation suggest that the flare QPP could be excited by these
recurrent jets.

Previous findings \citep[e.g.,][]{Reid12,Shen12,Lu19} found that
solar jets were always accompanied by solar flares, coronal
bright points, or filament eruptions. Recent observations also
showed that solar jets triggered by a solar flare had
repetitive and regular occurrences with a period of about 72~s, but
they did not find the similar quasi-period between flare
QPPs and recurrent jets \citep{Ning22}. The same quasi-period of
about 60~s was also discovered both in flare QPPs and recurrent
jets, and they took place almost simultaneously \citep{Shi22}.
However, it is impossible to conclude that whether these recurrent
jets have affected the flare QPP or they are just the result of the
flare QPP \citep[cf.][]{Ning22}. In our study, a same quasi-period
of 45~s is observed in both the flare QPP and recurrent jets, and
the onset time of these recurrent jets are $\sim$180-s
earlier than the beginning of flare QPP. Based on these
observational facts, we may infer that the flare QPP seen in the
SXR/HXR and microwave emissions is probably excited by recurrent
jets. The associated video (animation.mp4) shows that the
eruption of the first jet is like a mini-filament-driven jet very
much, indicating that the recurrent jets could be driven by the
eruption of mini-filaments that is associated with magnetic
reconnection \citep{Sterling20,Shen21}. Thus, both the recurrent
jets and the accompanying flare QPP could be associated with the
magnetic reconnection that is modulated by some periodic processes.

\subsection{Geometric and differential emission measure analysis}
\begin{figure}[h!]
\begin{center}
\includegraphics[width=0.8\linewidth]{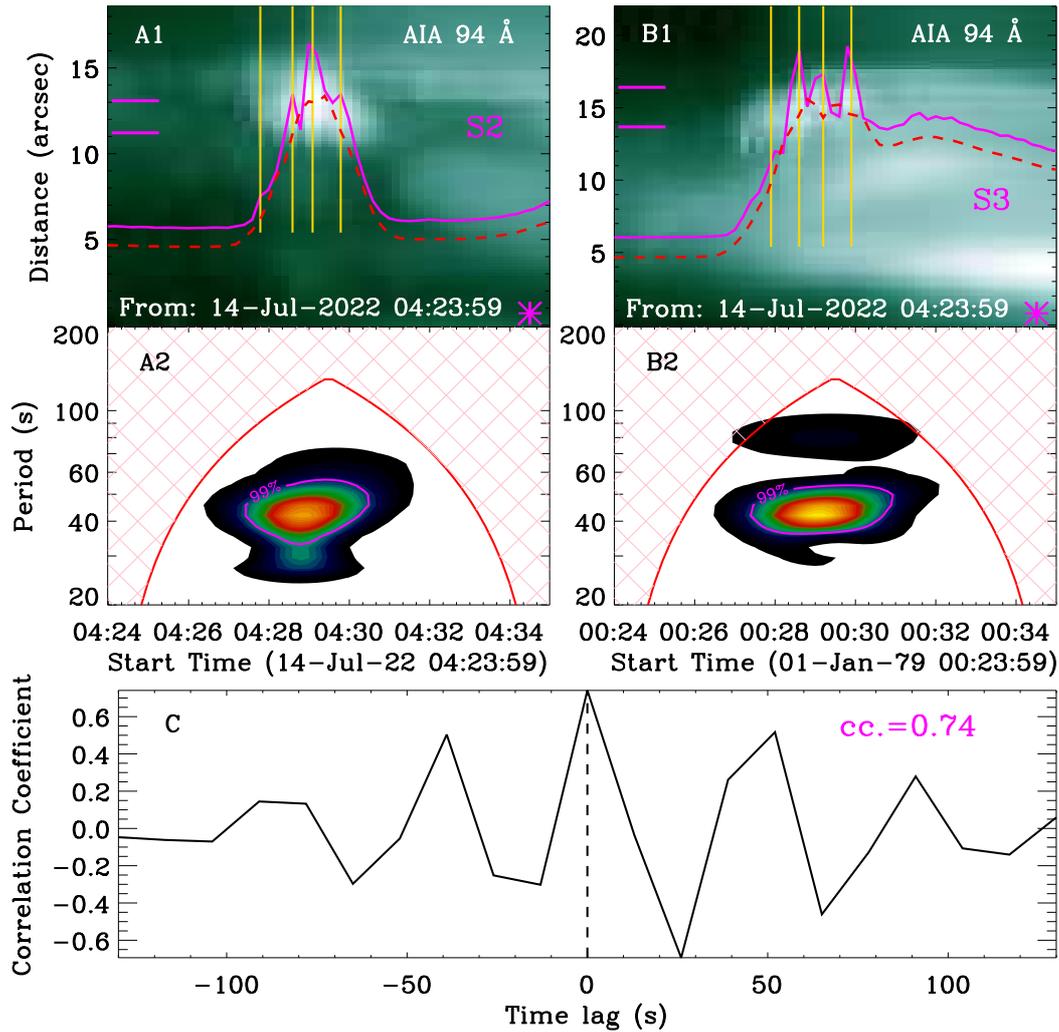}
\end{center}
\caption{The M1.2 flare observed by SDO/AIA~94~{\AA}. (A1-B1)
Time-distance diagrams along S1 and S2 (Figure~\ref{over}), the
magenta symbols ($\ast$) mark their start points. The overlaid solid
magenta curves are integrated from two short magenta lines on
the left hand, and the red dashed curves represent their slowly
varying components. (A2-B2) Morlet wavelet power spectra of the
correspond rapidly varying components. The magenta lines indicate
the significance level of 99\%. (C) Correlation coefficients between
two rapidly varying components as a function of the time lag, the
vertical line mark the time lag at 0~s.}\label{loop}
\end{figure}

\begin{figure}[h!]
\begin{center}
\includegraphics[width=0.8\linewidth]{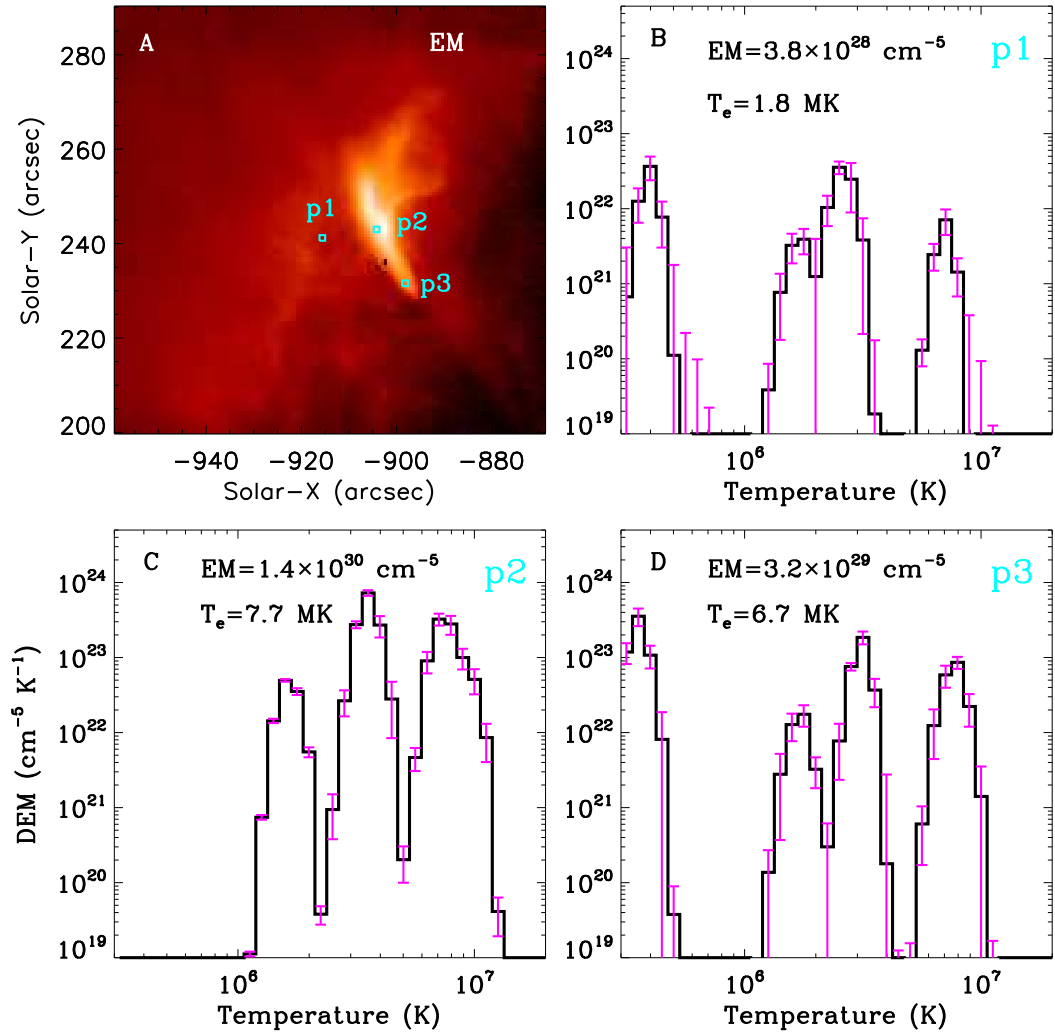}
\end{center}
\caption{DEM analysis results of the M1.2 flare. (A) EM map
integrated in the temperature range of 0.31$-$20~MK. Three cyan
boxes outline the non-flare region (p1), loop-top region (p2), and
footpoint (p3), respectively. (B-D) DEM profiles at the non-flare
region, loop top and footpoint. The EM and DEM-weighted average
temperature ($T_{\rm e}$) are also labeled in each
panel.}\label{dem}
\end{figure}

The flare QPP observed in SXR, HXR and microwave emissions could be
excited by a group of recurrent jets with the same intermittent
cadence, and they are most likely to be associated with a
nonthermal process, i.e., electron beams periodically
accelerated by the repetitive magnetic reconnection
\citep[e.g.,][]{Yuan19,Li21,Karampelas22}. In order to further know
whether the quasi-period of 45-s is modulated by an external MHD
wave \citep[e.g.,][]{Foullon05,Li15,Nakariakov18}, or it is only a
self-oscillating process \citep[e.g.,][]{Nakariakov10,Takasao16}, we
perform the geometric and differential emission measure (DEM)
analysis for the M1.2 flare, as shown in Figures~\ref{loop} and
\ref{dem}.

Figure~\ref{loop}~(A1-B1) present time-distance diagrams at
AIA~94~{\AA} along two slits (S2 and S3) in Figure~\ref{over}~(C),
and the magenta symbols (`$\ast$') mark their start points. Here,
the slits are selected to cross two opposite footpoints of
the flare loop, but they are not cross the loop top. Because there
are much more saturated pixels at loop top than those at footpoints
(see also the animation.mp4). In the two time-distance diagrams, it
does not see any signatures of displacement oscillations that are
perpendicular to loop legs. However, they appear clearly signatures
of brightness variations at double footpoints, as outlined by two
short magenta lines. Thus, the normalized light curves at
AIA~94~{\AA}, which are integrated intensities between two short
magenta lines, are overplotted in corresponding time-distance
diagrams, as shown by the solid magenta curves. Similar to the
microwave flux, at least four sub-peaks are found to superimpose on
the background emission, as indicated by the gold vertical lines,
which are less than those in HXR fluxes. They appear as
non-stationary QPPs, for instance, each pulsation is mainly
characterized by an anharmonic and triangular shape
\citep[e.g.,][]{Nakariakov19}. Using the same smooth window of 60~s,
the slowly (dashed red) and rapidly varying components are
distinguished from the raw light curves. Panels~(A2-B2) show Morlet
wavelet power spectra of the rapidly varying components at
AIA~94~{\AA}. They both reveal an enhanced power at the period
center of about 45~s from around 04:27:50~UT to 04:30:05~UT,
suggesting a dominant period of $\sim$45~s, similarly to what
observed in SXR/HXR and microwave emissions. Panel~(C) presents the
cross-correlation analysis \citep[e.g.,][]{Tian16} between two
rapidly varying components in AIA~94~{\AA} at double footpoints, the
maximum correlation coefficient of 0.74 is seen at the time lag of
0~s, as indicated by the vertical line. This observational result
suggest that the flare QPP at double footpoints is in phase.

Figure~\ref{dem} shows the DEM analysis result. It is calculated
from six EUV-wavelength observations measured by SDO/AIA. The
DEM($T$) distribution for each pixel is estimated by an improved
sparse-inversion code \citep{Cheung15,Su18}, and the DEM($T$)
uncertainty can be estimated from 100 Monte Carlo (MC) simulations,
for instance, the three times of standard deviations of 100~MC
simulations (3$\delta$). Panel~(A) presents the EM map integrated in
the temperature range of 0.31$-$20~MK. Similar to the
AIA~94~{\AA}~map, a post-flare loop can be seen in the EM map. Then,
three small regions (cyan boxes) with a FOV of about
1.8$''$$\times$1.8$''$ are selected to display DEM profiles, and
they are located at the non-flare region (or coronal background,
p1), loop-top region (p2), and one footpoint (p3), respectively.
Panels~(B-D) draw DEM profiles as the function of temperature, and
the error bars represent their uncertainties, i.e., 3$\delta$. The
EM and DEM-weighted mean temperature ($T_{\rm e}$) are calculated in
the temperature range between 0.31$-$20~MK, as labeled in each
panel. It can be seen that both the EM and $T_e$ at the loop top are
higher than that at the footpoint, and thus loop-top region is more
saturated. The $T_{\rm e}$ is estimated to $\sim$7.7~MK (C) at the
loop top and $\sim$6.7~MK (D) at the footpoint, which is consistent
with that the post-flare loop is most visible at AIA~94~{\AA}
(T$\approx$6.3~MK). At the non-flare region, the $T_{\rm e}$ is
$\sim$1.8~MK (B), which is roughly equal to quiet coronal
temperature.

\subsection{MHD explanation and Coronal seismology}
Based on the AIA~94~{\AA} map and EM map in Figures~\ref{over}~(C)
and \ref{dem}~(A), the distance between two footpoints of the flare
loop is estimated to $\sim$20.3~Mm, which leads to a loop length
($L$) of $\sim$31.9~Mm when assuming a semi-circular shape for the
flare loop \citep[cf.][]{Tian16,Gao22,Li22}. Then under the
assumption that the oscillation is associated with a standing wave,
the phase speed ($v_{\rm ph}$) can be determined by
Equation~(\ref{eq1}), for instance, twice the ratio of the loop
length to the quasi-period ($P$), which is about 1420~km~s$^{-1}$.

\begin{equation}
  v_{\rm ph} = \frac{2L}{P}.
\label{eq1}
\end{equation}

The local sound speed in the flare loop can be estimated by using
$v_{\rm s}\approx152\sqrt{T_{\rm e}/{\rm MK}}$
\citep[cf.][]{Nakariakov01,Kumar15,Li17b}. The average temperatures
at the loop top and footpoint are estimated to 7.7~MK and 6.7~MK,
which lead to the local sound speeds of $\sim$420~km~s$^{-1}$ and
$\sim$390~km~s$^{-1}$, respectively. Obviously, the estimated phase
speed of the flare loop is much faster than the local sound speeds
at the loop top and footpoints. Therefore, the 45-s period observed
in the M1.2 flare could not be modulated by the slow-mode wave in
the flare loop \citep[e.g.,][]{Wang21}, although the quasi-periods
less than 1~minute have been reported in flare QPPs and explained as
standing slow-mode waves \citep[e.g.,][]{Welsh06,Cho16}.

The estimated phase speed is much slower than that requires for the
global sausage-mode wave, i.e., the speed in the range of
$\sim$2400$-$5000~km~s$^{-1}$
\citep[e.g.][]{Nakariakov03,Melnikov05,Tian16}. Moreover, the global
sausage-mode wave is often found in the broader and denser plasma
loop, and the necessary condition is given by \cite{Nakariakov03} as
in Equation~(\ref{eq2}).

\begin{equation}
  \frac{n_{\rm i}}{n_{\rm o}} > (\frac{L}{0.65w})^{2}.
   \label{eq2}
\end{equation}
\noindent Here, $n_{\rm i}$ and $n_{\rm o}$ are the number densities
inside and outside of the flare loop (or non-flare region). $w$
stands for the loop width, and could regard as the full width at
half maximum of a Gaussian profile across the flare loop, which is
about 2.5~Mm. Thus, the density contrasty should be as high as 385
if the 45-s QPP is modulated by the global sausage-mode wave of
flare loop. The number density ($n_{\rm i}$) inside the flare loop
can be estimated by $\sqrt{{\rm EM}/w}$, which are
$\sim$7.5$\times$10$^{10}$~cm$^{-3}$ at the loop top and
$\sim$3.6$\times$10$^{10}$~cm$^{-3}$ at the footpoint. At the
non-flare region that has not plasma loops, the effective
line-of-sight depth (i.e., $w \approx 4 \times 10^{10}$~cm) is used
to calculate the $n_o$ \citep[see,][]{Zucca14,Li18,Suw18}, leading
to $\sim$9.7$\times$10$^{8}$~cm$^{-3}$. Then, the density contrast
is in the range of $\sim$37$-$77 from double footpoints to the loop
top. Such density contrast is rather low, compared to the necessary
condition of the global sausage oscillation in flare loops
\citep[e.g.,][]{Nakariakov03,Chen15}. Therefore, the
quasi-period at about 45~s seen in the M1.2 flare is impossible to
be modulated by the global sausage-mode wave of the flare loop.

In our study, the phase speed is quite close to the average speed of
about 1328~km~s$^{-1}$ in a catalog of kink-mode oscillations
\citep{Nechaeva19,Nakariakov21}, which are often identified as
transverse oscillations of plasma loops
\citep[e.g.,][]{Nakariakov99,Anfinogentov15,Suw18,Li20b,Tiwari21}.
In the corona, kink oscillations are always compressive, or
weakly compressive in the long wavelength regime
\citep{Goossens12,Nakariakov21}. On the other hand, they could be
seen as the brightness variation or intensity disturbance if the
loop displacement is not exactly perpendicular to the line-of-sight
\citep{Cooper03,Tian12,Wang12,Zimovets15,Antolin17,Li18}. In such
case, the local Alfv\'{e}n speed ($v_{\rm A}$) could be determined
by the phase speed ($v_{\rm ph}$) and the density contrast ($n_{\rm
o}/n_{\rm i}$), and the magnetic field strength ($B$) can be
estimated by using the local Alfv\'{e}n speed and mass density at
the loop top and footpoints, as shown in Equations~\ref{eq3} and
\ref{eq4} \citep[e.g.,][]{Yang20,Zimovets21,Tan22,Zhang22}.

\begin{equation}
  v_{\rm A} = v_{\rm ph}~(\frac{2}{1+n_{\rm o}/n_{\rm i}})^{-\frac{1}{2}}.
\label{eq3}
\end{equation}
%%%%%%%%%%%%%%%%%%%%%%%%%%%%%%%%%%%%%%%%%%%%%%%%%%%%%%%%%%%%%%%%%%%
\begin{equation}
  B \approx v_{\rm A}~(\mu_{\rm 0}~n_{\rm i}~m_{\rm p}~\widetilde{\mu})^{\frac{1}{2}}.
\label{eq4}
\end{equation}
\noindent Where, $\mu_{\rm 0}$ and $m_{\rm p}$ stand for the
magnetic permittivity of free space and the Proton mass, $n_i$ is
the number density at the flare loop, and $\widetilde{\mu} \approx
1.27$ represents the average molecular weight in the solar corona
\citep[e.g.,][]{Nakariakov01,Zhang20}. Then, the mass density
($\rho_{\rm i}$) could be roughly equal to $n_{\rm i}~m_{\rm
p}~\widetilde{\mu}$. Herein, the Alfv\'{e}n speed inside the
oscillating loop is estimated to about 1010~km~s$^{-1}$, leading to
the magnetic field strength of about 99~G and 143~G at the footpoint
and loop top, respectively. These strengths at the flare loop are
consistent with previous estimations in solar flares
\citep[e.g.,][]{Qiu09,Li17b,Li18,Zimovets21b}. Our measurement and
estimations support the idea that the quasi-period of 45~s in the
M1.2 flare could be modulated by the kink-mode wave of a flare loop
\citep{Nakariakov21}.

\section{Summary}
Based on observations recorded by Fermi, GECAM, GOES, SDO/EVE, and
NoRP, we investigate the non-stationary QPP at wavelengths of HXR,
SXR, microwave and EUV during the impulsive phase of an M1.2 flare
on 2022 July 14. Combined with the imaging observation from SDO/AIA,
the excitation and modulation of the flare QPP are discussed.
Our conclusions are summarized as following:

\begin{enumerate}

\item A quasi-period of $\sim$45$\pm$10~s is simultaneously detected at
Fermi~11.5$-$102.4~keV, GECAM~25$-$120~keV, GOES~1$-$8~{\AA},
ESP~1$-$70~{\AA}, NoRP~2~GHz and 3.75~GHz during the flare impulsive
phase, i.e., from about 04:27:50~UT to 04:32:20~UT. Our observations
suggest the coexistence of nonthermal and thermal processes in the
M1.2 flare, and the 45-s QPP at multiple wavelengths could share the
same periodic process of energy release, like the repetitive
magnetic reconnection \citep[e.g.,][]{Yuan19,Li21,Karampelas22}.

\item A group of recurrent jets with a periodicity of about
$\sim$45$\pm$10 are seen in AIA~304~{\AA} image series during
$\sim$04:24:50$-$04:32:20~UT. The onset time of the flare QPP is 180-s
later than that of recurrent jets, but they show the same
quasi-period, indicating that the flare QPP is probably excited by recurrent
jets. This observational result is different from previous findings,
for instance, solar jets were always triggered by the flare eruption
\citep{Reid12,Lu19}, or the periodicity of the solar flare and
accompanied jets is different \citep[e.g.,][]{Ning22}.

\item Thanks to the imaging observation from SDO/AIA at 94~{\AA}, the
quasi-period of $\sim$45$\pm$10~s is also seen at two
opposite footpoints of the flare loop. And the phase speed is
estimated to about 1420~km~s$^{-1}$. Our measurements imply that the
45-s period is most likely to be modulated by the kink-mode wave
\citep[cf.][]{Nakariakov10a,Nechaeva19}.

\item Based on the kink oscillation model, the Alfv\'{e}n speed inside the
flare loop is estimated to $\sim$1010~km~s$^{-1}$. The magnetic
field strengths are measured in the range of 99$-$143~G from the
footpoint to the loop top, similarly to what have estimated in solar
flares at the magnitude order of 100~G
\citep[e.g.,][]{Qiu09,Li18,Zimovets21b}.

\end{enumerate}

\section*{Conflict of Interest Statement}
The authors declare that the research was conducted in the absence
of any commercial or financial relationships that could be construed
as a potential conflict of interest.

\section*{Author Contributions}
D.~Li selected the topic, performed the main data analysis, led to
discussions and prepared the manuscript. F.~Shi contributed the
SDO/AIA data analysis and joined to modify the manuscript. H.~Zhao,
S.~Xiong, L.~Song, W.~Peng, and X.~Li provided the GECAM data
analysis. W. Chen contributed to analyse the Fermi data. Z.~Ning
joined to discuss the explanation of the flare QPP and recurrent
jets.

\section*{Funding}
This work is funded by the NSFC under grants 11973092, U1931138,
12073081, U1938102, as well as CAS Strategic Pioneer Program on
Space Science, Grant No. XDA15052200, and XDA15320301. D. Li is also
supported by the Surface Project of Jiangsu Province (BK20211402).

\section*{Acknowledgments}
The authors would like to acknowledge two anonymous referees
for their inspiring and valuable comments. We thank the teams of
Fermi, GECAM, GOES, SDO/AIA, SDO/EVE, and NoRP for their open data
use policy. GECAM is a mission funded by the Chinese Academy of
Sciences (CAS) under the Strategic Priority Research Program on
Space Science. SDO is a mission of NASA's Living With a Star Program
(LWS). AIA and EVE data are courtesy of the NASA/SDO science teams.
NoRP is operated by Solar Science Observatory, a branch of National
Astronomical Observatory of Japan, and their observing data are
verified scientifically by the consortium for NoRP scientific
operations.

\section*{Data Availability Statement}
The datasets for this study can be found here:
\href{https://fermi.gsfc.nasa.gov/ssc/data/}{https://fermi.gsfc.nasa.gov/ssc/data/},
\href{http://jsoc.stanford.edu/ajax/lookdata.html}{http://jsoc.stanford.edu/ajax/lookdata.html},
\href{https://lasp.colorado.edu/home/eve/data/}{https://lasp.colorado.edu/home/eve/data/},
\href{https://solar.nro.nao.ac.jp/norp/index.html}{https://solar.nro.nao.ac.jp/norp/index.html}.

%%Please see the availability of data guidelines for more information, at
%%https://www.frontiersin.org/about/author-guidelines#AvailabilityofData

\bibliographystyle{frontiersinSCNS_ENG_HUMS} % for Science, Engineering and Humanities and Social Sciences articles
\bibliography{my_refer}

\section*{Table captions}
\begin{center}
\captionof{table}{\label{tab1} Observational instruments/telescopes used in this work.}
\hspace{1.0em}
\begin{tabular}{ccccccc}
\hline \hline
Instrument & Wavelength       & Time cadence   &  Description  & Pixel scale  & Observation  \\
\hline
GOES       & 1$-$8~{\AA}      &     1~s        &     SXR       &     -        &   1D         \\
\hline
SDO/EVE/ESP & 1$-$70~{\AA}    &   0.25~s       &     SXR       &     -        &   1D         \\
\hline
Fermi/GBM  & 11.5$-$26.6~keV  & $\sim$0.256~s  &    SXR/HXR    &     -        &   1D         \\
           & 26.6$-$102.4~keV & $\sim$0.256~s  &     HXR       &     -        &   1D         \\
\hline
GECAM      & 25$-$120~keV     &   0.5~s        &     HXR       &     -        &   1D         \\
\hline
NoRP       &   2~GHz          &   1~s          &    radio      &     -        &   1D         \\
           &   3.75~GHz       &   1~s          &    radio      &     -        &   1D         \\
\hline
SDO/AIA    &  304~{\AA}       &  12~s          &    EUV        &    0.6$''$   &   2D         \\
           &   94~{\AA}       &  12~s          &    EUV        &    0.6$''$   &   2D         \\
\hline \hline
\end{tabular}
\end{center}

%%%%%%%%%%%%%%%%%%%%%%%%%%%%%%%%%%%%%%%%%%%%%%%%%%%%%%%%%%%%%%%%%%%%%
\begin{center}
\captionof{table}{\label{tab2} List of GRD numbers and their angles used in this work.}
\hspace{1.0em}
\begin{tabular}{ccccccccccccccc}
\hline \hline
GRD number          &  1    &    2    &     3    &     6     &   7     &    8    &    9    &    10   &   16  &    17   &   18    &   19   &   25      \\
\hline
Angle ($^{\circ}$)  & 72.9  &   43.5  &    72.7  &    82.6   &   52.2  &   31.2  &   52.0  &   79.1  &  64.5 &   39.4  &  34.2   &   62.8 &  49.8    \\
\hline \hline
\end{tabular}
\end{center}
NOTE--The angle refers to the incident angle of the GECAM/GRD from
the Sun.

\section*{Figure captions}

\end{document}